\documentclass[aip,apl
,amsmath,
amssymb
,reprint
]{revtex4-1}
\usepackage{graphicx}
\usepackage{dcolumn}
\usepackage[latin9]{inputenc}%für ß und ä,ö,ü
\usepackage{datetime} 
\usepackage{pifont} 
\usepackage{color}
\usepackage{natbib}
\begin{document}

%\preprint{AIP/123-QED}
\title{Mirrors for slow neutrons from holographic nanoparticle-polymer free-standing film-gratings}

\author{J. Klepp} 
\email[]{juergen.klepp@univie.ac.at}
\homepage[]{http://fun.univie.ac.at} 
\affiliation{University of Vienna, 
Faculty of Physics, 1090 Wien, Austria}
\author{C. Pruner}
\affiliation{University of Salzburg, 
Department of Materials Science and Physics, 5020 Salzburg, Austria}
\author{Y. Tomita}
\affiliation{University of Electro-Communications, 
Department of Engineering Science, 1-5-1 Chofugaoka, Chofu, Tokyo 182, Japan} 
\author{K. Mitsube}
\affiliation{University of Electro-Communications, 
Department of Engineering Science, 1-5-1 Chofugaoka, Chofu, Tokyo 182, Japan}
\author{P. Geltenbort} 
\affiliation{Institut Laue Langevin, 
Bo\^{i}te Postale 156, F-38042 Grenoble Cedex 9, France}
\author{M. Fally} 
\affiliation{University of Vienna, 
Faculty of Physics, 1090 Wien, Austria}
\date{\today}

\begin{abstract} 
We report on successful tests of holographically arranged grating-structures in nanoparticle-polymer composites in the form of 100 microns thin free-standing films, i.e. without sample containers or covers that could cause unwanted absorption/incoherent scattering of very-cold neutrons. Despite their large diameter of 2 cm, the flexible materials are of high optical quality and yield mirror-like reflectivity of about 90\% for neutrons of 4.1 nm wavelength.
\end{abstract}
\keywords{Neutron optics, Nanoparticles in polymers, Holographic gratings}

\maketitle

Today, neutron experiments are among the key techniques for materials science\cite{FurrerBook2009,WillisBook2009} as well as fundamental physics\cite{RauchWerner2000,HasegawaNJP2011,DubbersRMP2011}.  
In addition to well-established methods, further development of neutron spectrometers at existing and projected neutron research facilities are essential for progress in neutron physics. 
To advance such development, the design of new, efficient neutron-optical devices is necessary. 
Here, we report successful tests of holographic nanoparticle-polymer gratings with large area of about 2cm diameter. In particular, we have tested 100 microns thin free-standing film grating-structures, as recently developed for light-optics\cite{VitaAPL2007}. The gratings exhibit mirror-like behavior, i.e. a reflectivity of 0.9 for neutrons of wavelength $\lambda=4.1$ nm.   

Neutron optics is based on the one-particle Schrödinger equation, which contains the neutron-optical potential, i.e. the neutron refractive index at a certain wavelength\cite{Sears-89}.
The refractive index of a (nonmagnetic) material can be written as $n_0= 1-\lambda^2\, b_c\rho/(2\pi)$, in which $b_c$ is the coherent scattering length for a particular isotope and $\rho$ is the atomic number density of the material. 
A one-dimensional sinusoidal grating is
characterized by the periodically modulated refractive index $n(x)= n_0+\Delta n\cos(2\pi x/\Lambda)$, with modulation amplitude $\Delta n=\lambda^2~ b_c\Delta\rho/(2\pi)$ and grating spacing $\Lambda$. The quantity $\Delta\rho$ is the number-density modulation-amplitude.
Such grating structures can be produced by exploiting the light-induced change of the refractive index for neutrons in materials -- the photo-neutron-refractive effect\cite{Fally-apb02}. 
Using a conventional holography setup, coherent light beams
are superposed at the position of a recording material. Upon illumination with the resulting spatial light pattern a sinusoidal modulation of $b_c\rho$ and, therefore, also of the refractive index for neutrons occurs in the recording material.  
Diffraction of neutrons by thick holographic volume phase-gratings can -- analogically to neutron diffraction by thick crystals -- be described by dynamical diffraction theory\cite{FurrerBook2009,WillisBook2009,RauchWerner2000}. 
The reflectivity for neutrons in the symmetric Laue-case (transmission geometry) is written as 
\begin{eqnarray}\label{eq:Reflectivity}
R(\mbox{x},\mbox{y})
=\frac{\sin^2(\mbox{y}\sqrt{\mbox{x}^2+1})}{\mbox{x}^2+1},
\end{eqnarray} 
with 
\begin{eqnarray}\label{eq:x}
\mbox{x}
=\frac{2\pi\cos\theta_B}
{\lambda\,\Lambda\,b_c\Delta\rho}(\theta_B-\theta)\,\,\mbox{and}
\,\,\mbox{y}=\frac{\lambda\,d\, b_c\Delta\rho}{2\cos\theta}
\end{eqnarray} 
adapted for our requirements\cite{KleppNIMA2011}. Here, $d$, $\theta$ and $\theta_B$ are the thickness of the grating, the angle of incidence and the Bragg angle (as defined by $\lambda=2\Lambda\sin\theta_B$), respectively. 
Note that Eq.\,(\ref{eq:Reflectivity}) is equivalent to Kogelnik's theory for diffraction of light by thick hologram gratings\cite{KogelnikBellSysJ1969,KleppNIMA2011}. 
Now, the goal is to adjust tunable parameters such as $b_c$, $\Delta\rho$, $\Lambda$ and $d$, so that values of $R$ suitable for a particular application are reached. 
Holographic gratings could be used for various purposes in neutron optics, such as monochromators, polarizers, mirrors or beam splitters for matter-wave interferometry\cite{RauchWerner2000,HasegawaNJP2011,Pruner-nima06} and small-angle scattering applications with neutrons.

\begin{figure}
    \scalebox{0.38}
{\includegraphics {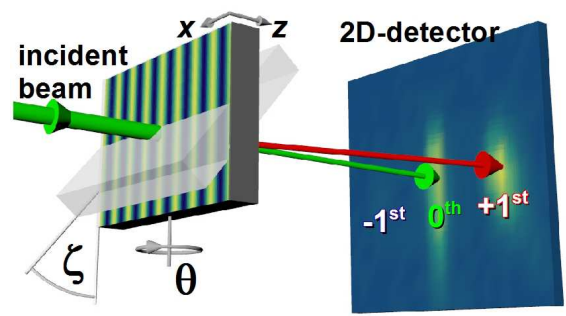}}
    \caption{Sketch of the experimental setup.}
    \label{fig1}
\end{figure}
Inorganic nanoparticles embedded in a photopolymer matrix (nanoparticle-polymer composites) have already been investigated intensively \cite{SuzukiAPL2002,SuzukiApplOpt2004,SakhnoNanotech2007} for light-optics applications.
In nanoparticle-polymer gratings, the refractive-index modulation $\Delta n$ for neutrons can be tuned by choosing a suitable value of $b_c$, i.e. the species of nanoparticles. 
The thickness of the recorded gratings -- and thereby the neutron reflectivity -- is, so far, limited to about 100 microns by detrimental light scattering during recording. This obstacle can be overcome by exploiting the so-called Pendell\"{o}sung interference effect\cite{FurrerBook2009,WillisBook2009,RauchWerner2000}: Due to the quantum-mechanical superposition of several neutron waves formed in the periodic potential of crystals or holographic gratings, the neutron intensity is swapped back and forth between diffracted and forward-diffracted (transmitted) beams. This effect depends on the incident neutron wavelength\cite{Shull-prl68} and the effective thickness\cite{Sippel-pl65} of the sample [cf. Eq.\,(\ref{eq:x})]. The latter can be increased by tilting\cite{SomenkovSolStComm1978} holographic gratings of limited thickness around an axis parallel to the grating vector. Therefore, the Pendell\"{o}sung interference effect allows to enhance the reflectivity via tilting. 
Only recently, the feasibility of a beam splitter for cold neutrons has been demonstrated by such means\cite{FallyPRL2010,KleppPRA2011}. 
Up to now, the main limitations of neutron-optical elements based on nanoparticle-polymer composites have been the small grating diameter -- decreasing the possible beam size upon tilting -- and incoherent scattering/absorption of sample covers and the sample itself. 

The SiO$_2$ nanoparticles used for the present investigation have an average core diameter of about 13\,nm\cite{SuzukiApplOpt2004}. They are produced by liquid-phase synthesis and dissolved in methyl isobutyl ketone solution. The SiO$_2$ sol is dispersed to multifunctional methacrylate monomer. The nanoparticle concentration was 20 vol\%.
As radical photoinitiator, 1 wt.\% titanocene (Irgacure784, Ciba) is added to enable the monomer to photo-polymerize at wavelengths shorter than 550 nm. The mixed syrup is cast on a glass plate and is dried. 
Spacers are arranged around the sample before it is covered with another glass plate to obtain film samples of about 100\,$\mu$m thickness and diameter larger than 2 cm, about double of what had been achieved before.
At this stage of the preparation, the photoinitiator, the monomer and the nanoparticles are homogeneously distributed in the sample material. 
Next, two expanded ($\approx 2$ cm in diameter), mutually coherent and $s$-polarized laser beams of equal intensities at a wavelength of 532 nm are superposed to create a spatial sinusoidal light-intensity pattern at the sample position. The resultant grating spacing was $\Lambda=0.5\,\mu$m.
Via the photoinitiator, the pattern induces polymerization in the bright sample regions, a process that consumes monomers. As a consequence of the growing polymer-concentration gradient, 
nanoparticles counter-diffuse from bright to dark regions\cite{TomitaOptLett2005}, resulting in an approximately sinusoidal nanoparticle-concentration 
pattern. Subsequent homogeneous illumination ensures that the material is fully polymerized so that the nanoparticle density-modulation remains stable for years. Now, to obtain free-standing film gratings of only 100 microns thickness, both glass plates had been silanized beforehand and can be separated from the film samples by carefully cutting along the sample edges with a razor blade after recording. The samples are, to a certain extent, elastic and of excellent homogeneity across the whole sample area as is demonstrated by the neutron experiments explained below.

Neutron diffraction by such gratings was investigated using the very-cold neutron (VCN) beam at PF2 of the Institut Laue Langevin (ILL) in Grenoble, France.
The measurement principle is sketched in Fig.\,\ref{fig1}. 
The gratings are mounted in transmission geometry. Tilting the gratings to the angle $\zeta$ around an axis parallel to the grating vector -- in order to adjust the effective thickness -- $\theta$ was varied to measure rocking curves in the vicinity of the Bragg angle $\theta_B$. 
The neutron wavelength distribution of the incident beam is very broad, typically more than $\Delta\lambda/\lambda\approx 30$\%.
The collimation-slit width and distance were chosen so that the beam divergence was below 1 mrad. 
Experimentally, the reflectivity for the $i^{\mbox{\scriptsize{th}}}$ diffraction order is obtained as 
\begin{eqnarray}\label{eq:diffrEff}
R_{i}=I_{i}/I_{\mbox{\scriptsize{tot}}},
\end{eqnarray}
where $I_i$ and $I_{\mbox{\scriptsize{tot}}}$ are the measured intensities of the $i^{\mbox{\scriptsize{th}}}$ diffraction order and the total intensity of transmitted and diffracted beams, respectively. For each $\theta$, the sum over all 2D-detector pixels ($^3$He multi-wire proportional chamber, position resolution: 2$\times$2 mm$^2$, efficiency: 80\% at $\lambda=4$ nm) in each separated spot -- associated to one single diffraction order -- was calculated and the obtained intensities (corrected for background) plugged into Eq.\,(\ref{eq:diffrEff}).
\begin{figure}
    \scalebox{0.44}
{\includegraphics {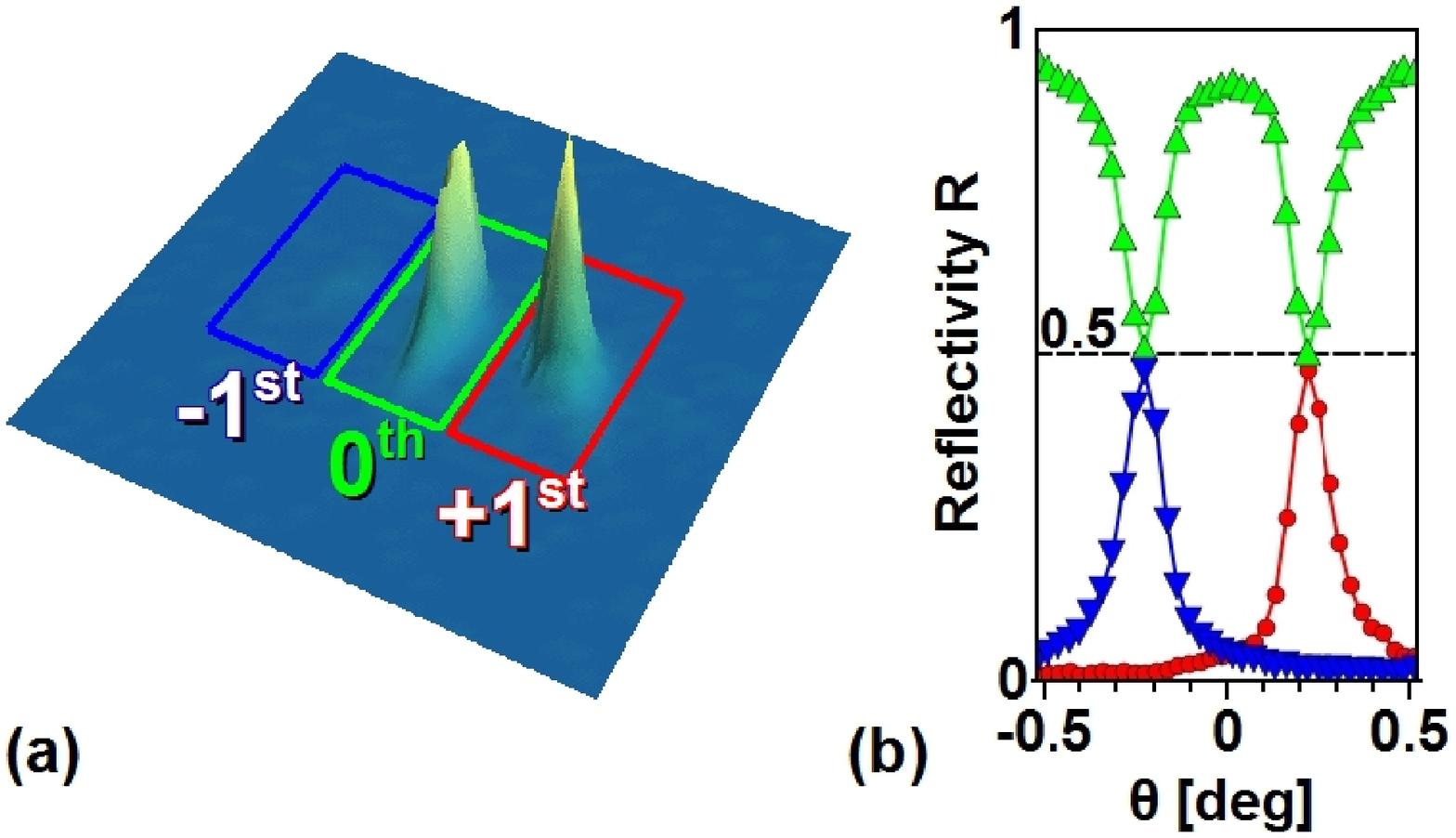}}
    \caption{(a) 2D-detector matrix at $\theta=\theta_B$. (b) Reflectivity calculated from Eq.\,(\ref{eq:diffrEff}) (\textcolor{blue}{\ding{116}}
$\dots -1^{\mbox{\scriptsize{st}}}$ order, \textcolor{green}{\ding{115}}
$\dots 0^{\mbox{\scriptsize{th}}}$ order,  \textcolor{red}{\ding{108}}
$\dots +1^{\mbox{\scriptsize{st}}}$ order). The grating is effectively a beam splitter.}
    \label{fig2}
\end{figure} 
\begin{figure*}
    \scalebox{0.77}
{\includegraphics {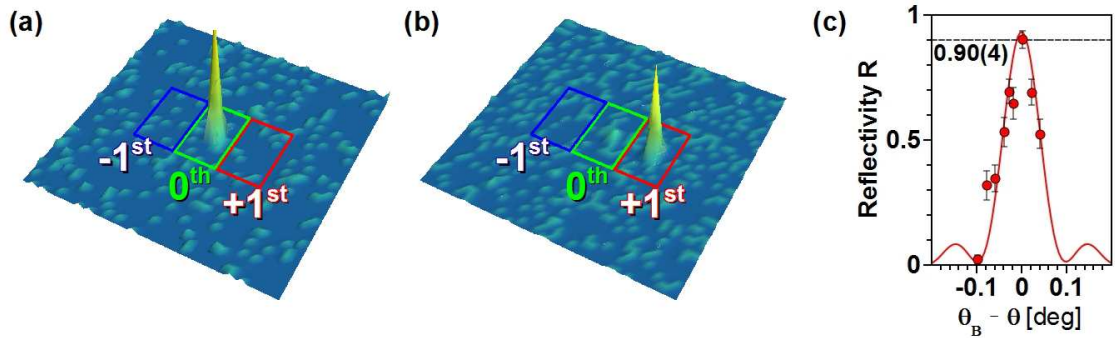}}
    \caption{(a) 2D-detector matrix for an off-Bragg position. (b) 2D-detector matrix for $\theta=\theta_B$. (c) Reflectivity for the $+ 1^{\mbox{\scriptsize{st}}}$ diffraction order as calculated from Eq.\,(\ref{eq:diffrEff}). The solid line is a theory curve according to Eq.\,(\ref{eq:Reflectivity}), taking into account the width of the wavelength distribution and the beam divergence of about 1mrad. The peak value of $R=0.90(4)$ corresponds to the data shown in (b). The grating exhibits mirror-like reflectivity.}
    \label{fig3}
\end{figure*}
Images of the 2D-detector matrix at $\theta=\theta_B$ and a rocking curve for the tilt angle $\zeta= 67^\circ$ are shown in Figs.\,\ref{fig2}(a) and \ref{fig2}(b), respectively. 
Here, the full incident spectrum with peak wavelength of $\lambda_0\approx 4$ nm was used. 
Note that no $-1^{\mbox{\scriptsize{st}}}$ order diffraction-peak is observed in Fig.\,\ref{fig2}(a). This means that the assumption of two-wave coupling -- necessary to arrive at Eq.\,(\ref{eq:Reflectivity}) -- is fulfilled. Furthermore, one can clearly see that the incident beam is transformed into two beams of equal intensity, their propagation directions enclosing an angle of $2\theta_B=0.47^\circ$. In this configuration, the free-standing film-grating acts as a beam splitter 
since $R_{\pm1}$ is approximately 0.5 as shown in Fig.\,\ref{fig2}(b). Measurements comparing glass-covered samples and free-standing films show that intensity loss due to absorption/incoherent scattering by glass covers is of the order of a few percent. In this regard, the advantage of free-standing films will -- because the absorption cross section is proportional to $\lambda$ -- become significant at larger wavelengths used for VCN interferometry\cite{VanDerZouwNIMA2000} ($\lambda\approx 10$ nm). Furthermore, since the film gratings are flexible, they should be well-suited for producing bent focusing neutron monochromators\cite{mikulaJApplCryst1986}.

Next, to demonstrate mirror-like reflectivity, the tilt angle is increased to about $70^\circ$. Moreover, the incident spectrum must be narrowed. Otherwise the rocking curve broadens and the observable peak-reflectivity is reduced. We achieved a wavelength distribution as narrow as $\Delta\lambda/\lambda\approx 1$\% by measuring the time of flight (TOF) of the neutrons on the distance from a disk-chopper -- inserted before the sample -- to the detector position and post-selecting only those that arrive in a narrow time-interval of 0.5 ms. Since the chopper we used has a duty cycle (open-to-closed ratio) of about 3\%, measurement times are extended from minutes or hours up to several days. In future experiments, this difficulty can be overcome by replacing the TOF-system by a multilayer mirror with distributed layer thickness, as tested in Ref.\,\onlinecite{HoghojJPhysSocJp1996} 
and successfully applied for VCN interferometry\cite{VanDerZouwNIMA2000} 
at $\Delta\lambda/\lambda\approx 10$\%. 
Next, $\theta$ was varied in TOF-mode to find the $+ 1^{\mbox{\scriptsize{st}}}$ diffraction order peak position, where the intensities of the diffraction spots were measured for several days. 
Raw data for $\theta$ far from $\theta_B$ (off-Bragg) and for $\theta=\theta_B$ is shown in Figs.\,\ref{fig3}(a) and (b), respectively. One can see that the pronounced peak of the transmitted beam at the center of Fig.\,\ref{fig3}(a) has almost vanished in Fig.\,\ref{fig3}(b), while here the intensity is at maximum at the position of the 
$+1^{\mbox{\scriptsize{st}}}$ diffraction order peak. 
In Fig.\,\ref{fig3}(c) a plot of the reflectivity for $\lambda=4.1$ nm is shown. The peak reflectivity is as high as 0.90 . Clearly, the film-grating acts as a mirror for $\theta=\theta_B$. 

To conclude, we have demonstrated the implementation of flexible free-standing holographic gratings of large diameter and only 100 $\mu$m thickness as beam splitters and mirrors for VCN. Such diffractive optical elements are anticipated to play an important role in the next generation of neutron interferometers as well as in small-angle scattering applications for VCN\cite{YamadaNIMA2011}.

We thank M. Yamada (Kyoto University) and Y. Iwashita (Institute of Chemical research, Kyoto University) for leaving their TOF system at our disposal. Also, we truly appreciate provision of the detector by G. Manzin and B. Guerard (ILL). Furthermore, we are deeply indebted to T. Brenner (ILL) for outstanding technical support and his readiness to lend a helping hand. Financial support by the Austrian Science Fund (FWF): P-20265, and the Ministry of Education,
Culture, Sports, Science and Technology of Japan (Grant
No. 23656045) is greatly acknowledged.

%\bibliographystyle{C:/Users/juergen/data/LaTex/localtexmf/bibtex/bst/apsrev}
%\bibliography{C:/Users/juergen/data/LaTex/localtexmf/bibtex/bib/juergen}

\end{document}